# CONSTRAINTS ON THE FORMATION HISTORY OF THE ELLIPTICAL GALAXY NGC 3923 FROM THE COLORS OF ITS GLOBULAR CLUSTERS

Stephen E. Zepf [1]

*Department of Astronomy, University of California*

*Berkeley, CA 94720*

*e-mail: zepf@astron.berkeley.edu*

Keith M. Ashman

*Department of Physics and Astronomy, University of Kansas*

*Lawrence, KS 66045*

*e-mail: ashman@kusmos.phsx.ukans.edu*

Doug Geisler

*Cerro Tololo Inter-American Observatory*

*National Optical Astronomy Observatories* [2]

*Casilla 603, La Serena, Chile*

*e-mail: dgeisler@ctio.noao.edu*

## Abstract

We present a study of the colors of globular clusters associated with the elliptical galaxy NGC 3923. Our final sample consists of Washington system $C$ and $T_1$ photometry for 143 globular cluster candidates with an expected contamination of no more than 10%. We find that the color distribution of the NGC 3923 globular cluster system (GCS) is broad and appears to have at least two peaks. A mixture modeling analysis of the color distribution indicates that a two component model is favored over a single component one at a high level of confidence ($> 99\%$). This evidence for more than one population in the GCS of NGC 3923 is similar to that previously noted for the four other elliptical galaxies for which similar data have been published. Furthermore, we find that the NGC 3923 GCS is redder than the GCSs of previously studied elliptical galaxies of similar luminosity. The median metallicity inferred from our $(C - T_1)_0$ colors is $[\text{Fe/H}]_{med} = -0.56$, with an uncertainty of 0.14 dex arising from all sources of uncertainty in the mean color. This is more metal rich than the median metallicity found for the GCS of M87 using the same method, $[\text{Fe/H}]_{med} = -0.94$. Since M87 is more luminous than NGC 3923, this result points to significant scatter about any trend of higher GCS metallicity with increasing galaxy luminosity. We also show that there is a color gradient in the NGC 3923 GCS corresponding to about -0.5 dex in $\Delta[\text{Fe/H}]/\Delta \log r$. We conclude that the shape of the color distribution of individual GCSs and the variation in mean color among the GCSs of ellipticals are difficult to understand if elliptical galaxies are formed in a single protogalactic collapse. Models in which ellipticals and their globular clusters are formed in more than one event, such as a merger scenario, are more successful in accounting for these observations.

## 1. Introduction

The origin of elliptical galaxies is a critical question for cosmological models and theories of galaxy formation and evolution. One approach has been to take at face value the old age and relatively uniform stellar populations of elliptical galaxies and model their formation as the collapse of isolated, massive objects at early epochs ($z > 2$). Models of the dissipational collapse of such systems can account for some of the basic properties of elliptical galaxies (Larson 1975, Carlberg 1984). Furthermore, Arimoto & Yoshii (1987) and Matteuci & Tornambè (1987) have shown that a model of star formation, chemical evolution, and galactic winds in an isolated, massive gas cloud can reproduce the color-luminosity relationship of elliptical galaxies. However, the lack of rotation observed in bright elliptical galaxies, combined with their high central densities, is difficult to understand if they formed in a monolithic dissapational collapse. Such models can only be saved by an efficient mechanism for transfering angular momentum from the visible central regions to large radii (cf. Silk & Wyse 1993). Dissipational collapse models also tend to predict steep color gradients which increase with luminosity, contrary to observational evidence (e.g. Franx, Illingworth, & Heckman 1989, Peletier et al. 1990).

An alternative approach is to consider models in which ellipticals are built from mergers, as originally postulated by Toomre & Toomre (1972). This scenario accounts for the low specific angular momentum of elliptical galaxies much more naturally than single collapse models. Moreover, N-body simulations of the mergers of spirals produce remnants which generally resemble elliptical galaxies (e.g. Barnes & Hernquist 1992). The merging of galaxies is also expected in at least a general way from cosmological models in which structure is built through hierarchical clustering. Detailed studies have demonstrated the importance of merging to galaxy evolution in specific hierachical clustering scenarios such as Cold Dark Matter (Cole et al. 1994; Kauffmann, White, & Guiderdoni 1994; and references therein).

Observationally, the merger hypothesis is supported by studies of nearby merging systems which find that elliptical galaxies *can* form from the merger of spiral galaxies (e.g. Wright et al. 1990, Schweizer 1990). However, the frequency of major mergers at the current epoch appears to be lower than that required to make all elliptical galaxies this way in a Hubble time (Toomre 1977). This argument does not hold if merging was much more common in the past, as suggested on theoretical grounds by Carlberg (1990), Toomre (1977), and others. In addition, several observational studies have reported tentative detections of a higher frequency of interactions or mergers in the past (Zepf & Koo 1989; Burkey et al. 1994; Carlberg, Pritchet, & Infante 1994; Colless et al. 1994). A similar increase in the frequency of interaction or mergers also appears to be observed in studies of galaxies in clusters (e.g. Couch et al. 1994; Dressler et al. 1994; Lavery, Pierce, & McClure 1992). Thus the merger hypothesis remains plausible but unproven.

The properties of the globular cluster systems provide a valuable tool for distinguishing between competing models of the formation of elliptical galaxies. In the merger scenario, at least two populations of globular clusters around ellipticals are expected, one population associated with the progenitor spirals, and one formed during the merger itself (Ashman & Zepf 1992). The formation of globular clusters in interactions and mergers, originally predicted by Ashman & Zepf (1992) and Schweizer (1987), now appears to be confirmed by HST observations of NGC 1275 (Holtzman et al. 1992), NGC 7252 (Whitmore et al. 1993), and He 2-10 (Conti & Vacca 1994). The population of globular clusters formed during the merger will be younger, more metal-rich, and more spatially concentrated than the low metallicity, spatially extended halo population of the progenitor spirals. In most cases, the difference in metallicity between the two populations is expected to be observable as a significant color difference. Age differences are also expected, but usually both populations will be rather old because most mergers occur at moderately high redshifts. In this case, models of stellar populations (e.g. Worthey 1994) clearly predict that the metallicity differences will dominate the optical colors. Therefore, one signature of a merger origin is a GCS color distribution with at least two peaks - one which is blue and metal-poor, and another which is red and relatively metal-rich.

Monolithic collapse models of elliptical galaxy formation predict a very different GCS color distribution than the merger model. Generically, simple models of a monolithic collapse give metallicity distributions with a smooth shape and a single peak (e.g. Arimoto & Yoshii 1987). Within the context of specific models, the age and metallicity distribution of the objects can be described in detail and thus compared directly to observations. In all of these pictures, elliptical galaxies are assumed to form in a single event at an early epoch, so the color distribution is expected to be simply related to the metallicity distribution.

The color distribution of the GCS of an elliptical galaxy can thus provide critical evidence for determining the formation history of the galaxy. Although the scenarios we describe for the merger and monolithic collapse models are unquestionably simplistic, we expect that the general trends will hold. For example, peaks in the color distribution clearly point to episodic formation of some type, which would require fundamental changes in the monolithic collapse picture, but are natural in the merger model. Conversely, a smooth, single-peaked color distribution is natural in the monlithic collapse picture, but would require contrived circumstances in general in a merger scenario.

The most effective way to address the question of the color distributions of GCSs is through imaging over a large area in both the blue and the red or near-infrared in order to obtain large samples with high metallicity sensitivity. Until the recent advent of large format CCDs with high quantum efficiency in the blue, such studies were time-consuming and therefore rare. Zepf & Ashman (1993) analyzed the color distributions of two elliptical galaxies, NGC 4472 and NGC 5128. These were chosen because they had published photometry in the metallicity sensitive $B-I$ and $C-T_1$ indices (Couture et al. 1991 for NGC 4472, Harris et al. 1992 for NGC 5128). Zepf & Ashman (1993) found that the color distribution of the NGC 4472 GCS was better fit by a two component distribution than a single component one at a 98.5% confidence level. They found a similar preference for bimodality in the NGC 5128 GCS at a 95% confidence level. Ajhar, Blakeslee, & Tonry (1994) present photometry of a significant number of globular cluster candidates in the less metallicity sensitive $V - I$ index for five ellipticals in the Virgo cluster. Our quantitative analysis of these data confirms the bimodality of the NGC 4472 GCS, and finds that the GCS color distributions of two additional ellipticals have strong bimodal signatures, while two others do not. In addition, three cD galaxies, NGC 1399 (Ostrov, Geisler, & Forte 1992), M87 (Lee & Geisler 1993), and NGC 3311 (Secker et al. 1994) have also been analyzed in this way. The color distribution of the GCSs of these cD galaxies appears to be complex, with the unimodal hypothesis being ruled out at a high confidence level (97% to 99%).

The primary goal of this paper is to study the color distribution of the GCS of NGC 3923 in order to derive constraints on the formation history of this galaxy. We present the observations and data analysis in Section 2. We then consider the distribution of color and compare it to competing models of elliptical galaxy formation in Section 3. In Section 4, we consider the median color of the NGC 3923 GCS, and compare it to that of other elliptical galaxy GCSs. In Section 5, the color gradient of the NGC 3923 GCS is determined. We discuss these results and the implications for elliptical galaxy formation in Section 6.

## 2. Observations and Data Reduction

We present a study of the GCS of NGC 3923, a bright elliptical galaxy with an absolute magnitude of $M_B = -21.1$, based on $B_T^0 = 10.5$ and the $D_n - \sigma$ distance of 1587 km s$^{-1}$ from Faber et al. (1989), with $H_0 = 75$ km s$^{-1}$ Mpc$^{-1}$. NGC 3923 is located in a loose group of galaxies. It is most famous for its shells, a common feature of ellipticals located outside of galaxy clusters (Schweizer & Seitzer 1992). In all other respects, such as galaxy color, metallicity, and velocity dispersion, NGC 3923 appears to be an ordinary elliptical galaxy.

In order to determine the colors of the GCS of NGC 3923, we obtained deep images of a field centered on the galaxy with the 4m at CTIO on UT 20 Feb 1993. With the TEK1024 # 1 CCD at prime focus, the image scale was $0.46''$ per pixel. We obtained integrations of $7 \times 1000$s in $C$ and $5 \times 1000$s in $T_1$, where $C$ and $T_1$ are passbands in the Washington photometric system (Canterna 1976). This system is useful for our purpose since it combines wide bandpasses necessary for observing faint objects with a broad wavelength baseline for estimating metallicities (Geisler & Forte 1990, hereafter GF). These authors derive a metallicity calibration for the $(C - T_1)_0$ color of

$$[\text{Fe/H}] = 2.35(C - T_1)_0 - 4.39$$

based on the study by Harris & Canterna (1977) of the integrated colors of Galactic globular clusters, ranging in [Fe/H] from $-2.25$ to $-0.25$. A comparison to a similar calibration of $(B - V)$ and $(V - I)$ indicates that $(C - T_1)$ is about twice as sensitive to metallicity differences at a given photometric precision.

The individual images of the NGC 3923 field were processed in the standard way within IRAF, and then registered and median combined. Stellar images in the combined frames have a FWHM of $\sim 1.5''$. We performed photometry on the objects in the field using DAOPHOT as implemented in IRAF. The background light of the galaxy was removed through iterative median filtering (e.g. Fischer et al. 1990). The point spread function (psf) was defined by $\sim 20$ bright, unsaturated, and uncrowded stars throughout the field. The psf of these stars varies with position on the chip, with a roughly linear increase of about 20% going from the sharpest stars in the southeast corner to the broadest in the northwest. Therefore, we utilized the variable psf feature in DAOPHOT. For the $C$ frame, we found no measureable deviation from the fitted linearly variable psf, and the conversion to large aperture magnitudes was constant over the whole field. However, a slight residual variation after variable psf-fitting was detected in the $T_1$ frame. This is manifested in a conversion to large aperture magnitudes which varies slightly across the field. After a small correction for this effect, the random uncertainty introduced to the $T_1$ photometry by these slight deviations from the fitted psf is only $0.01 - 0.02$ mag. This is insignificant compared to other sources of photometric uncertainty.

We also examined the image shapes of the objects classified by DAOPHOT in order to distinguish extended objects from point-like sources. This step is useful since GCs are point sources at the distance of NGC 3923, but background galaxies are generally extended in images of this depth and seeing (e.g. Harris et al. 1991). We employed a combination of the image shape estimators in the DAOPHOT package and those described by Harris et al. (1991) to separate background galaxies from our GC sample. Our procedure and classification limits are described in more detail in Zepf, Geisler, & Ashman (1994b, hereafter ZGA2)

The instrumental magnitudes were then transformed to standard $C$ and $T_1$ magnitudes. Since conditions during the 4m observations were not photometric, the deep images were calibrated by observations of the NGC 3923 field in photometric conditions with the 0.9m at CTIO on UT 21 March 1993. Observations of 44 Washington system standards (Harris & Canterna 1979, Geisler 1990 and 1994) were obtained on this night. The rms errors of these standards were 0.015 in $T_1$ and 0.030 in $(C - T_1)$, indicating the night was of good photometric quality. The zero point and color term of the transformation to standard $T_1$ and $(C - T_1)$ magnitudes for the deep 4m data were determined by photometry of 12 uncrowded stars in the NGC 3923 field which were measured on both the 4m and 0.9m frames. The resulting transformations are

$$T_1 = t_1 - 0.035(c - t_1) - 3.27$$

$$(C - T_1) = 1.262(c - t_1) - 0.26$$

with an rms of 0.017 and 0.024 respectively.

These resulting magnitudes and colors were dereddened, using $E(B - V) = 0.06$ (Burstein & Heiles 1982), $A_{T_1} = 2.62 E(B - V)$, and $E(C - T_1) = 1.97 E(B - V)$ (Geisler 1994). The internal uncertainties for colors of individual objects are approximately 0.04 mag at $(T_1)_0 = 21.0$ and 0.08 mag at $(T_1)_0 = 22.0$. We note that comparisons to external $(C - T_1)_0$ values must include the uncertainties in the reddening and the photometric calibration. A complete list of the photometry is given in ZGA2.

We present the results of this photometry in Figure 1, where we plot the $(C - T_1)_0$, $(T_1)_0$ color-magnitude diagram. In this diagram, the GCS of NGC 3923 is clearly evident as a large abundance of objects with $(T_1)_0 \gtrsim 20.4$, and $1.2 \lesssim (C - T_1)_0 \lesssim 2.1$. Because of incompleteness, the number of detected objects begins to decline by $(T_1)_0 = 22.5$, and there are few objects fainter than 23.0. For reference, the peak or "turnover" of the GC luminosity function is roughly at $(T_1)_0 = 23.8$ (Zepf, Geisler, & Ashman 1994a), and the color of the galaxy itself is $(C - T_1)_0 = 1.87$ (see section 4).

Lines of constant incompleteness at the 10% and 50% level are also shown in Figure 1. These were determined by extensive artificial star tests, which are described in more detail in ZGA2. From these tests, we determined the completeness of the sample over the entire range of $(C - T_1)_0$, $(T_1)_0$ to a precision of 1-2%. One notable feature is that the incompleteness depends on color, with a higher completeness for blue objects compared to red ones at a given $(T_1)_0$ magnitude. This is a result of our combined $C$ image being shallower than our $T_1$ image, and is also influenced by the substantial color term in the transformation to the standard $(C - T_1)_0$ color. Future observations will be much better in this regard since the Tek 2048 CCDs now available have about twice the sensitivity in $C$ as the Tek 1024 we used.

In order to create from this data set a sample of globular cluster candidates which is as large as possible, but not severely affected by incompleteness or contamination, we adopt as our primary sample the 143 point sources with $20.4 < (T_1)_0 < 22.1$, $1.05 < (C - T_1)_0 < 2.15$, and which are within the elliptical area $55.2'' < \sqrt{(ab)} < 230.0''$, where the ellipticity is 0.4 and $a$ and $b$ are the semi-major and semi-minor axes respectively. Considering each of these limits in turn, the magnitude limit is set at the bright end by the onset of the NGC 3923 GCS. The faint limit is set where the incompleteness of the extreme red limit of the sample reaches 50%. The completeness for typical objects in the sample is then very high, with a median value of about 90%.

Even with these conservative magnitude limits, an unbiased determination of properties such as the median color or shape of the color distribution must account for the color dependence of the incompleteness. The principle on which we base our correction for this effect is that the probability that any given object is detected is directly proportional to the completeness at its given magnitude and color. Thus we can create an unbiased sample by weighting each object by the inverse of the completeness at its point in the color-magnitude diagram. We implement this idea in a Monte Carlo fashion, by assigning the object in the sample least likely to be detected a weight of one, and then randomly sampling the rest of the population, including objects with a probability equal to the ratio of their completeness to that of the first object. We repeat this process many times in order to achieve a fair sampling of the distribution. We emphasize that the corrections are small. However, we use this procedure to insure that our results are not affected by a color related bias.

The color limit, $1.05 < (C-T_1)_0 < 2.15$, is designed to eliminate foreground stars from the sample, while excluding only a minimal number of globular clusters. This technique works because the majority of the foreground stars are dwarfs in the Galactic disk, which are much redder than any globular cluster and fall well outside of these limits. At the same time, it is clear from Figure 1 that the GCS of NGC 3923 falls inside of these limits with few, if any, globular clusters lost as a result.

We also impose radial limits on our sample. The inner radial cutoff is determined by the point at which the procedure to remove the background light from the galaxy breaks down and the photometry of the objects begins to be affected. The outer radial cutoff is a tradeoff between trying to maximize the sample size without including significant contamination. At our adopted outer radial limit, the surface density of the NGC 3923 GCS is roughly 6 times greater than our best estimate of the background. However, if we use our upper limit to the background density, then the overdensity of objects associated with NGC 3923 drops to 1.5. Thus our adopted outer radial limit is a compromise choice to give a reasonable sample size for the most likely background estimate while not overly compromising the sample if the background level is closer to our upper limit. The radial limits and the ellipticity of the GCS are discussed in more detail in ZGA2.

Our expectation is that the sample we have defined is composed almost exclusively of globular clusters associated with NGC 3923. The contamination from foreground stars within our sample limits predicted by the Bahcall & Soniera model of the Galaxy (Bahcall 1986), is 8.8 stars, or 6% of the total sample. Several background galaxies may have also slipped through our image classification procedure. Although this number is expected to be small (e.g. Harris et al. 1991), we can estimate an upper limit to contamination from background galaxies using studies of galaxy counts. This analysis gives an upper limit to the total contamination of the sample from both stars and galaxies of 19%. We note that this upper limit would require the unlikely classification of almost half of the background galaxies as stellar. Therefore, we conclude that our NGC 3923 sample is most likely composed of $\gtrsim 90\%$ GCs, in agreement with a similar study of NGC 4472 by Couture et al. (1991), which used similar selection criteria.

### 3. Color and Metallicity Distribution of the NGC 3923 GCS

The metallicity distribution of the GCS provides a valuable record of the formation history of the host galaxy. Since the integrated color of old stellar systems is determined primarily by metallicity, broadband colors are an observationally efficient way to determine the metallicity distribution of a GCS. Color indices with good sensitivity to metallicity variations, such as the $(C-T_1)_0$ color of the Washington photometric system, are particularly useful in this regard. Therefore, we use $(C-T_1)_0$ colors to study the metallicity distribution of the NGC 3923 GCS. In Figure 2, we present a histogram of the colors of our NGC 3923 globular cluster sample. Along the top axis, we also provide the metallicities corresponding to these colors using the relationship of GF between $(C-T_1)_0$ and [Fe/H] given in the previous section (noting that it is extrapolated to higher metallicities than the Galactic GCS on which it is based).

The color distribution in Figure 2 appears to have two separate peaks with a slight deficit between them. The bimodal appearance of the histogram suggests a formation history with at least two distinct episodes of globular cluster formation. The distribution is similar to that expected in a merger model, and is unlike the smooth, single-peaked distribution expected in a monolithic collapse model. In order to quantitatively assess the significance of the bimodality in the color distribution, we utilize the KMM mixture-modeling algorithm (McLachlan & Basford 1988). This algorithm objectively partitions a dataset into two (or more) components and then assesses the improvement of the two-group fit to the one-group fit using the likelihood ratio test statistic (see Ashman, Bird, & Zepf 1994 for a discussion of this algorithm in the context of astronomical applications). For the color distribution of the GCS of NGC 3923, the KMM algorithm rejects the unimodal hypothesis in favor of the bimodal one at a confidence level of $> 99\%$.

The rejection of the single component fit in favor of the two component fit at the $> 99\%$ level supports a merging process rather than a monolithic collapse as the primary formation mechanism of NGC 3923. This confidence level is the median value of 100 realizations of the sample, using the Monte Carlo technique described in the previous section to make the small correction for the color dependence of the incompleteness. This technique is very stable, with only modest deviations among the realizations. For example, 98% of the realizations reject the unimodal hypothesis at a confidence level of at least 90%. We also note that the "uncorrected" 143 object sample gives a similarly strong bimodal signal (> 99% confidence level). Morever, the general shape of the color distribution is independent of the choice of magnitude limits, which can be seen in the color-magnitude diagram shown as Figure 1.

The translation of the observed color bimodality into a metallicity bimodality also appears to be robust. If the linear relationship between color and metallicity of GF is adopted, then such a translation is trivial. Even with the somewhat non-linear behavior of the color-metallicity relationship suggested by theoretical models (e.g. Worthey 1994), the metallicity distribution of the NGC 3923 GCS is very poorly fit by any single component model, and appears to be composed of at least two populations. Furthermore, the evidence for two populations in the colors stands regardless of what is driving the color variations.

The two component fit to the NGC 3923 GCS found by the KMM algorithm partitions the sample into one component with 54% of the objects at a mean of $(C - T_1)_0 = 1.47$, and a second with 46% of the objects at a mean $(C - T_1)_0 = 1.87$. The mean metallicities of these two components are -0.94 and 0.00 respectively, based on the GF calibration. The wide separation between these two components highlights the broad color distribution of the NGC 3923 GCS. The shape of these individual components is assumed to be gaussian in our current implementation of the KMM algorithm (see discussion in Ashman et al. 1994). In the context of this study, such a shape seems reasonable given the good fit of a normal distribution to the metallicity distribution of the Milky Way halo GCS (Armandroff & Zinn 1988), the roughly normal shape of the metallicity distributions predicted by monolithic collapse models (e.g. Arimoto & Yoshii 1987), and the lack of evidence for any other form of the distribution.

It is very unlikely that the shape of the color distribution of our sample is significantly affected by contamination by objects which are not globular clusters associated with NGC 3923. This is because of the low level of background contamination in our sample and the smooth distribution in color of the likely sources of this contamination. Foreground stars are expected to make up about 6% of the total sample, based on the Galactic model of Bahcall & Soniera. Galactic models also predict a smooth color distribution of foreground stars within the color limits of our sample. Similarly, any background galaxies which have been misclassified as stars are likely to be smoothly distributed in color. The distribution of color among the objects which we classify as galaxies is smooth within the limits of our sample, in agreement with the color distributions found in faint galaxy surveys (Koo & Kron 1992 and references therein). Therefore, we conclude that the color distribution of our sample accurately reflects the colors of the globular clusters of NGC 3923, and that the distribution of these colors is significantly better fit by a mixture of two populations with normal distributions and different mean colors than by any single normally distributed population.

## 4. The Median Color of the NGC 3923 GCS

The average color of the GCS is interesting since it provides information about the overall extent of metal enrichment in the GCS. As reviewed by Harris (1991), it is well-established that the mean colors of the GCSs of elliptical galaxies are bluer and by inference more metal-poor on average than their host galaxies. Harris (1991) and Brodie & Huchra (1991) corroborate the general trend first noted by van den Bergh (1975) that GCS metallicity tends to be higher in more luminous galaxies. This trend can also be considered in terms of a correlation of GCS metallicity with host galaxy morphology, since the most luminous galaxies studied are ellipticals, with spirals at intermediate luminosity, and dwarfs the faintest.

For the NGC 3923 GCS, we find a mean $(C-T_1)_0$ color of 1.65, with a total uncertainty of 0.06. This corresponds to [Fe/H] = -0.51 based on the GF calibration. The median color is $(C-T_1)_0 = 1.63$, giving a median metallicity of [Fe/H] = -0.56. The uncertainty due to the color alone is 0.14 dex for these metallicities. Because the median is less sensitive to outliers and non-linearities in the metallicity calibration, we adopt it as our primary estimator of the average color and metallicity.

The derived median metallicity of [Fe/H] = -0.56 is higher than that reported for the GCS of any galaxy, with the exception of the GCS of NGC 3311, as communicated to us by Secker et al. (1994). A particularly good comparison can be made between the median NGC 3923 GCS color and the median GCS colors of the four other giant ellipticals (M87, NGC 1399, NGC 3311, and NGC 5128) which have been observed in the same $(C - T_1)_0$ color. Table 1 gives the absolute magnitude of the host galaxy, number of globular clusters in the sample, median $(C - T_1)_0$ color, and the inferred metallicity for each of these GCSs. Note that we have used the Burstein & Heiles (1982) reddening values for all of the galaxies, which are slightly different from the values used in some of the original papers. These data suggest that there is signficant scatter in the average GCS color and metallicity at a given magnitude for elliptical galaxies.

In order to specifically address the question of the signficance of the red color of the NGC 3923 GCS relative to previously published GCSs of elliptical galaxies like M87, we have carefully considered each source of uncertainty in the color and searched for any possible systematic effects. One test of the accuracy of the colors of our NGC 3923 GCS sample is a comparison of the colors we find the galaxy itself to published photometry. Since the color of the galaxy is well-known, this provides a valuable check on the zero point of our photometry. From our photometry, we derive a galaxy color of $(C - T_1) = 1.99 \pm 0.03$ within an aperture of $30''$. Using a conversion from $(C-T_1)$ to $(B-V)$ based on common standard stars, our photometry gives $(B - V) = 1.02 \pm 0.05$. This is in excellent agreement with the Faber et al. (1989) value of $(B - V) = 1.01 \pm 0.02$ within the same aperture.

A second check on the accuracy of our photometry is provided by frames of M87 taken during the same night used for calibration of the deep NGC 3923 images. Aperture photometry was obtained for a number of objects on these frames which have published photometry obtained from deep KPNO 4m images (Lee & Geisler 1993). The differences between the two sets of photometry are $\Delta(C - T_1) = 0.035 \pm 0.038$ (mean error) for 10 objects, and $\Delta T_1 = -0.039 \pm 0.035$ for 12 objects, where the differences are given as 0.9m - Lee & Geisler. Therefore, we find no significant evidence for errors in the photometry of the NGC 3923 GCS colors, and can rule out a zero point error at the 0.10 mag level or greater at a high confidence level.

The possibility that the reddening along the line of sight to NGC 3923 has been underestimated is strongly constrained by the tightness of the relationship between the $(B - V)_0$ color and $Mg_2$ stellar absorption line index (Burstein et al. 1988). The scatter about this relationship is observed to be only about 0.023 mag. Moreover, most of that small scatter is accounted for by the measurement uncertainties in $(B - V)$ and $Mg_2$. Thus, there is very little room left for errors in relative $E(B - V)$ values, which contribute additional scatter by adding uncertainty to the $(B - V)_0$ values while not affecting the $Mg_2$ determination.

Figure 3 is a plot of $(B - V)_0$ against $Mg_2$ for the elliptical galaxy sample of Faber et al. (1989), including the galaxies with $\log\sigma > 2.0$, $M_B < -17.0 + 5 \log h$, and those with a $(B - V)_0$ measurement (totalling 318 galaxies). The dashed line corresponds to $(B - V)_0 = 1.12\ Mg_2 + 0.615$, which is the relationship between these quantities found by Burstein et al. (1988). Clearly the scatter about this line is small, and NGC 3923 is well fit by the relationship with its standard reddening value. Specifically, the $(B - V)_0$ value of NGC 3923 predicted by its reddening independent $Mg_2$ value is 0.96, which is 0.01 redder than the observed $(B - V)_0$ of 0.95. If the $E(B - V)$ value was increased by 0.08 mag, as necessary to make the median color of the NGC 3923 GCS the same as that of M87, then the $(B - V)_0$ color of the galaxy would be 0.09 mag bluer than expected based on its reddening independent $Mg_2$ index. Ellipticals this much bluer than expected from their $Mg_2$ index are extremely rare. In this sample of ellipticals only 2 of the 318 galaxies are this blue or bluer for their $Mg_2$ index. A similar conclusion is reached if the velocity dispersion is used in place of the $Mg_2$ index. Thus, it is very unlikely that the red color of the NGC 3923 GCS is a result of underestimated reddening along the line of sight.

In conclusion, we are unable to find any other explanation for our observed colors other than that the NGC 3923 GCS is intrinsically red. Independent confirmation of this red color and inferred high metallicity would clearly be valuable. Spectroscopic absorption line indices would be particularly useful in this regard. However, the color alone supports the existence of significant scatter in the average GCS metallicity at a given elliptical galaxy magnitude.

## 5. The Color Gradient of the NGC 3923 GCS

Radial gradients in the average globular cluster metallicities also provide insight into the formation history of the GCS and the parent galaxy. Metallicity gradients can be particularly useful in constraining the amount of dissipation involved in the formation process. In Figure 4, we plot $(C-T_1)_0$ against log radius for our NGC 3923 GCS sample. This plot clearly shows the large color spread at all radii discussed in previous sections. In addition, there appears to be a small gradient such that clusters at larger radii are slightly bluer. Quantitatively, a linear least squares fit gives

$$(C-T_1)_0 = 1.72(\pm 0.03) - 0.21(\pm 0.08) \log r,$$

where $r = \sqrt{ab}$ in arcseconds. The Pearson correlation coefficient indicates that a negative color gradient is detected at 99.3% confidence. This fit, which is based on a radially extended sample, is not significantly changed if a more restricted radial range is adopted. The significance of the correlation and the values of the slope and zero point are also robust to changes in the limiting magnitude of the sample.

Thus, the NGC 3923 GCS displays a small but significant color gradient. In terms of metallicity, the change is roughly -0.5 dex in $\Delta[\text{Fe/H}]/\Delta \log r$. As noted previously, the GCSs of four other ellipticals have been studied using the same metallicity sensitive $(C - T_1)$ index we have used here. These studies also cover approximately the same radial range as our study (roughly 5-25 kpc). The GCSs of M87 (Lee & Geisler 1993), NGC 1399 (Ostrov et al. 1992), and NGC 3311 (Secker et al. 1994) appear to have gradients similar to the one we observe in the NGC 3923 GCS. No significant gradient was detected in the initial analysis of the NGC 5128 GCS (Harris et al. 1992). However, Zepf & Ashman (1993) have suggested that the two very blue clusters at small radii may be young rather than extremely metal-poor, in which case the NGC 5128 GCS would have a color gradient consistent with the other three GCSs. Although Ajhar et al. (1994) do not find gradients in $V - I$ to be common in their sample, the modest gradients observed in $C - T_1$ would be difficult to detect in $V - I$, which is only half as sensitive to metallicity variations. Therefore, based on the very small sample currently available, metallicity gradients may be the norm for the GCSs of elliptical galaxies.

The observed metallicity gradients place some constraints on the amount of dissipation involved in the formation of the GCS and the elliptical galaxy. Most models of elliptical galaxy formation involve dissipation in some way, either during a single collapse, or during a merger of spiral galaxies. For example, in the particular merger model of Ashman & Zepf (1992), color gradients in elliptical galaxy GCSs are a natural consequence of the composite nature of these systems and the greater spatial extent of the metal-poor clusters of the progenitor spirals. That some dissipation is involved in the formation of elliptical galaxies was already indicated by the existence of color and line-strength gradients in the integrated stellar light of most ellipticals (e.g. Davies, Sadler, & Peletier 1993 and references therein). Typical gradients for the integrated light of elliptical galaxies are roughly -0.2 dex in $\Delta[\text{Fe/H}]/\Delta \log r$, suggesting that the metallicity gradients observed in our very small sample of elliptical galaxy GCSs may be larger than those in the integrated light. However, the comparison is not direct since the stellar gradients are typically measured inside an effective radius, whereas the GCS gradients are measured at larger radii.

Discussion

The colors of the NGC 3923 GCS provide three critical constraints on the formation history of this elliptical galaxy. These are: 1) the distribution of color is fit much better by a bimodal distribution than a unimodal one, 2) the median color is red compared to the several other elliptical galaxies which have been studied in a similar way, and 3) there is a color gradient.

The first and second of these observations are inconsistent with a single collapse picture of the formation of elliptical galaxies. Such a picture naturally predicts a smooth, single-peaked distribution of color for the GCS, which is not observed. There is also no mechanism within this picture to account for the significant variation in median GCS color among elliptical galaxies of similar luminosity and velocity dispersion. In fact, monolithic dissipational collapse models predict an increase in mean GCS metallicity with elliptical galaxy luminosity, since the deeper potential wells of more luminous (massive) galaxies allow more recycling of gas through stars and hence higher metallicities (cf. Arimoto & Yoshii 1987, Brodie & Huchra 1991). Athough a single collapse model with dissipation can reproduce the color gradients observed within GCSs, the failures of this model indicate that a different picture of elliptical formation is required.

Our observations of the colors of the NGC 3923 GCS provide significant evidence that the formation history of elliptical galaxies includes several distinct episodes. In particular, Ashman & Zepf (1992) predicted that the metallicity distribution of elliptical galaxy GCSs would have two or more components if ellipticals form from the mergers of spirals. The observational confirmation of this prediction is a major success for merger scenarios, particularly since no other model has been able to account for more than one peak in the color distribution.

The variation in the average GCS metallicity of ellipticals of similar luminosity is also unsurprising if elliptical galaxies are formed from the mergers of spiral galaxies. Because of the stochastic nature of mergers and the composite nature of elliptical galaxy GCSs in Ashman-Zepf type models, several different processes may have an important effect on the average metallicity. Among the physical parameters which might (and are often expected to) vary among different mergers are - the ratio of globular clusters formed in the merger to those associated with the progenitor spirals, the metallicity of the globular clusters formed in mergers, the metallicity of the globular clusters associated with the progenitor spirals, especially if "thick-disk" populations are common, and the accretion of smaller satellites and their lower metallicity gas and clusters by the elliptical galaxy after its primary merger event.

It is beyond the scope of this paper to consider in detail the relative importance of each of these effects. However, the merger scenario offers promising possibilities for understanding the variation of average GCS color among ellipticals. This is in contrast to the serious difficulties monolithic collapse pictures have with this observation. As the number of elliptical galaxy GCSs with good color data increases, it will be possible to place tighter constraints on the relative importance of the various effects in determining the metallicities of GCSs around ellipticals.

A radial color gradient like that observed in the NGC 3923 GCS is also accounted for in a merger model because the higher metallicity clusters formed in the merger are more spatially concentrated than the metal-poor clusters associated with the progenitor spirals (Ashman & Zepf 1992). Taken as a whole, the data indicate that the formation process of elliptical galaxies and their globular clusters involves more than one episode and includes a stochastic element. The formation of ellipticals by merging spirals fits this description well.

We thank Bill Harris for making his image classification code available to us, Terry Bridges for comments on the manuscript, and Sidney van den Bergh for a helpful and stimulating referee's report. We also thank Jeff Secker and collaborators for communicating results in advance of publication and Ed Ajhar for providing tabular material in electronic form. S.E.Z. acknowledges support from NASA through grant number HF-1055.01-93A awarded by the Space Telescope Science Institute, which is operated by the Association of Universities for Research in Astronomy, Inc., for NASA under contract NAS5-26555. K.M.A. acknowledges support from a Fullam/Dudley Award and a Dunham Grant from the Fund for Astrophysical Research. K.M.A. thanks the Astronomy Department of the University of California, Berkeley for hospitality during the preparation of this paper, and the Center for Research, Inc., University of Kansas for travel support. This work has made use of the NASA/IPAC Extragalactic Database (NED), which is operated by the Jet Propulsion Laboratory, California Institute of Technology, under contract with the National Aeronautics and Space Administration.

**TABLE 1: Median $(C - T_1)_0$ of Elliptical Galaxy GCSs**

| Galaxy | $M_B$ | $Nobj$ | $(C - T_1)_0$ | [Fe/H] |
|---|---|---|---|---|
| NGC 3923 | -21.1 | 143 | 1.63 | -0.56 |
| NGC 1399 | -20.6 | 256 | 1.55 | -0.75 |
| NGC 3311 | -22.1 | 103 | 1.74 | -0.31 |
| M87 | -21.3 | 407 | 1.47 | -0.94 |
| NGC 5128 | -20.4 | 62 | 1.45 | -0.98 |

Notes - Col. (1) is the name of the host galaxy; col. (2) is the absolute $B$ magnitude of the host galaxy, with photometry from the RC3 for NGC 5128 and Faber et al. (1989) for the other galaxies, and distances from Faber et al. for NGC 3923 and NGC 3311 and from Ciardullo, Jacoby, & Tonry (1993) for NGC 1399, M87, and NGC 5128; col. (3) is the number of globular clusters from which the following two columns are determined; col. (4) is the median $(C - T_1)_0$ color of the GCS, taken from the sources given in the text; col. (5) is the metallicity corresponding to the median color using the GF calibration.

**Figure Captions**

Figure 1 - A $(C-T_1)_0$, $(T_1)_0$ color-magnitude diagram for all objects with stellar-like images within our radial limits in the NGC 3923 field. The GCS of NGC 3923 is clearly seen as an overabundance of objects with $(T_1)_0 \gtrsim 20.4$ and with color around $(C-T_1)_0 \approx 1.65$. The dotted line represents the 90% completeness limit, and dashed line is the 50% completeness limit.

Figure 2 - A histogram of the $(C-T_1)_0$ color distribution for the GCS of NGC 3923. The shape is clearly non-gaussian, and is fit much better by a bimodal distribution than a unimodal one. The Geisler & Forte (1990) metallicity calibration of the $(C-T_1)_0$ index is given along the top axis. This indicates that the GCS of NGC 3923 contains few very metal-poor clusters. The sample is defined as those point sources with $20.4 < (T_1)_0 < 22.1$, $1.05 < (C-T_1)_0 < 2.15$, and the histogram includes the small correction for the dependence of the completeness on color as described in the text. The bin size is 0.1 mag, about twice the typical internal error.

Figure 3 - A plot of $(B-V)_0$ color vs. $Mg_2$ absorption-line index for 318 early-type galaxies in the Faber et al. (1989) sample. NGC 3923, which is represented by the star, lies in the middle of the narrow locus of points. Any increase in the correction for reddening along the line of sight to NGC 3923 would cause it to lie blueward of its expected color given its reddening independent $Mg_2$ value.

Figure 4 - The plot of $(C-T_1)_0$ color against galactocentric radius for our NGC 3923 GCS. Note that here we show a more extended radial range ($46.0'' < sqrt(ab) < 414.0''$) in order to gain a longer radial baseline. The formal fit to the color gradient is shown as the dashed line, and the GF metallicity calibration of the color is given along the right hand axis.

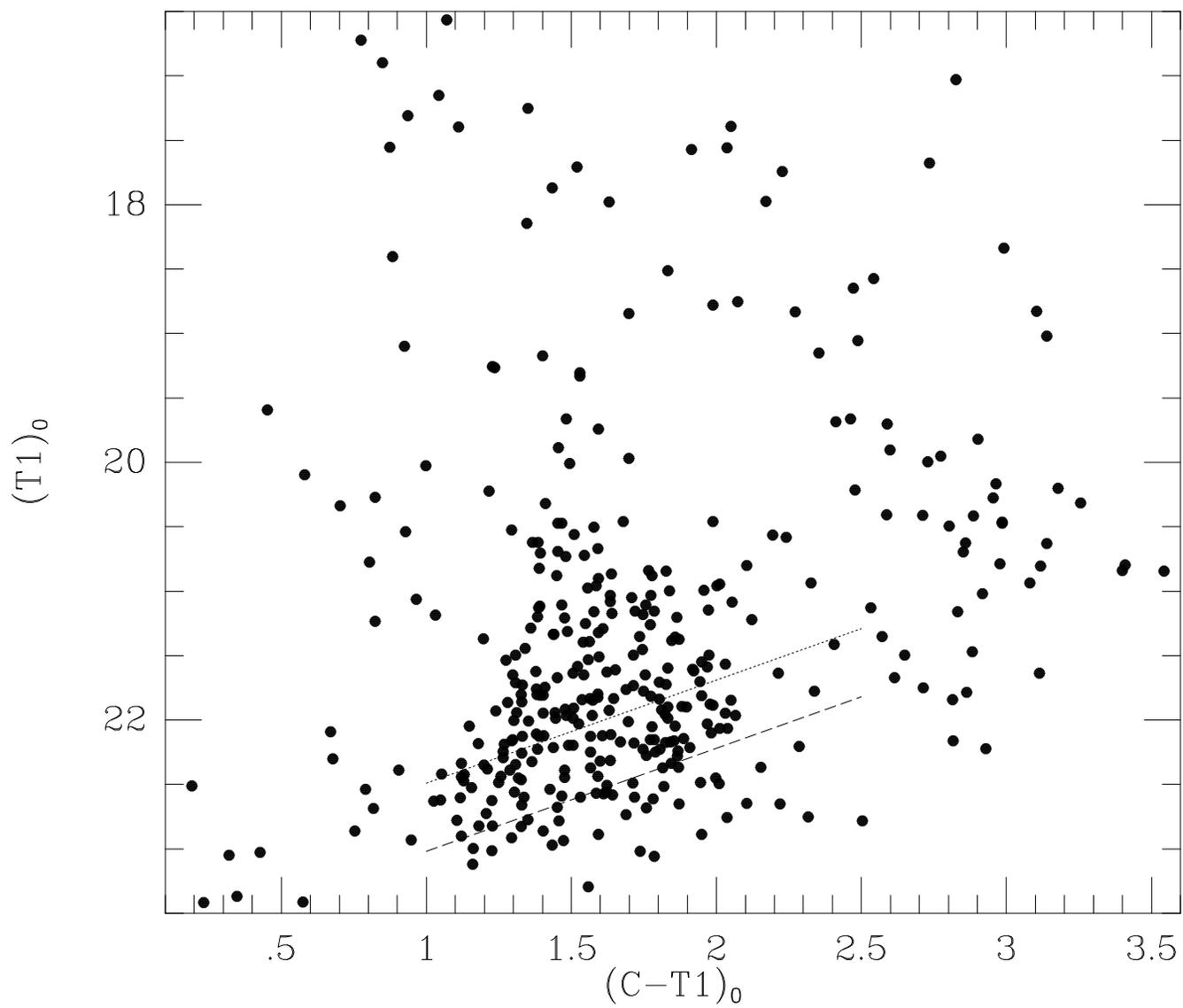

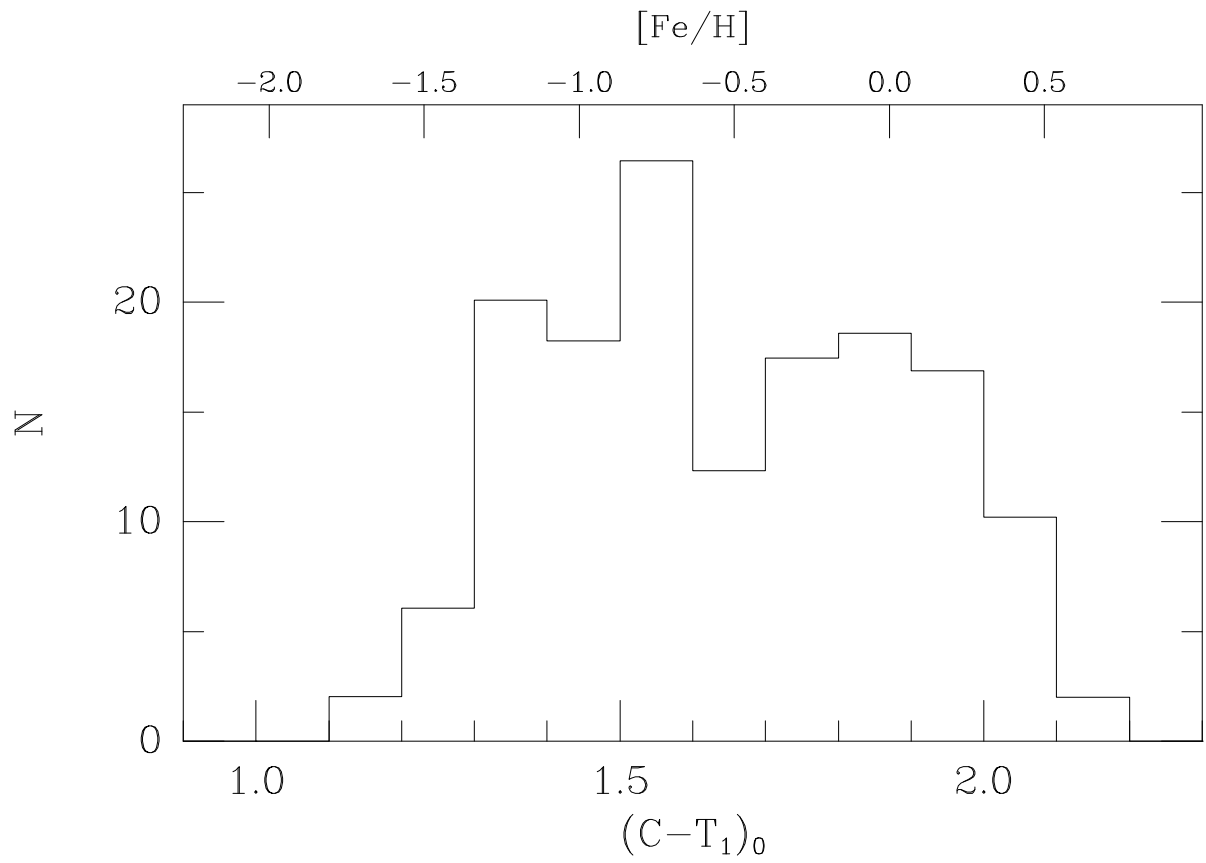

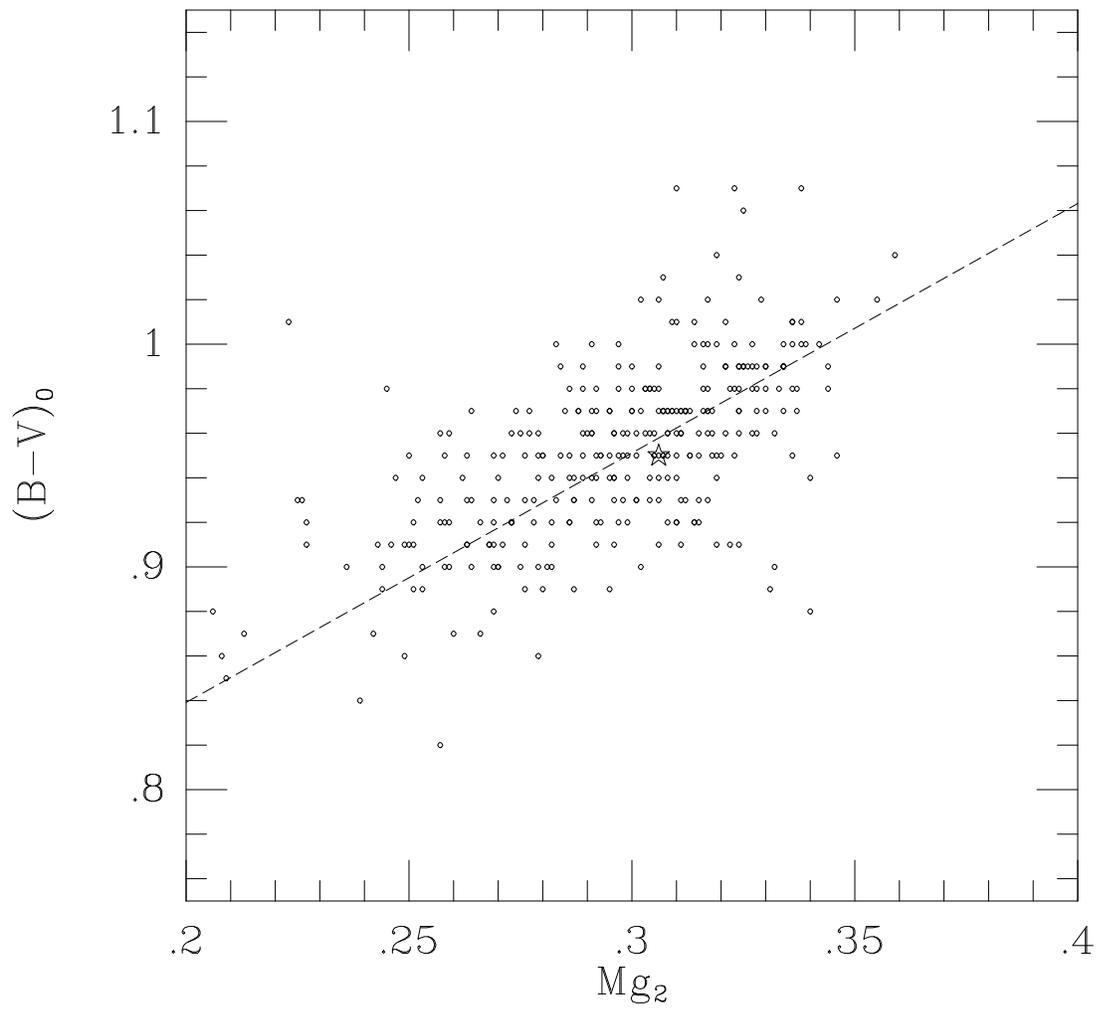

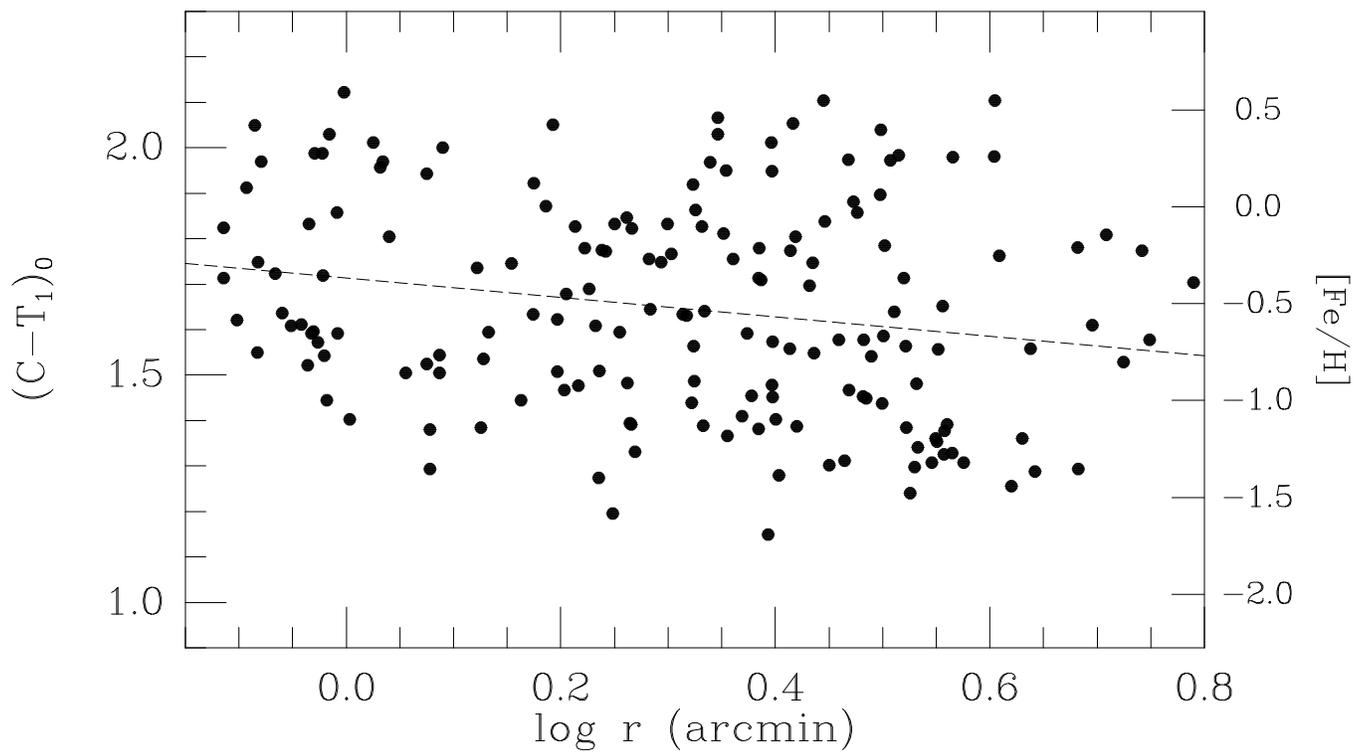